\documentclass[prd,twocolumn,superscriptaddress,showpacs,preprintnumbers,amsmath,amssymb,nofootinbib,APS]{revtex4}

\usepackage{dcolumn}
\usepackage{bm}
\usepackage[latin1]{inputenc}
\usepackage[spanish,english]{babel}
\usepackage{amsfonts}
\usepackage{amssymb}
\usepackage{graphicx}

\newcommand{\be}{\begin{equation}}
\newcommand{\ee}{\end{equation}}
\newcommand{\bea}{\begin{eqnarray}}
\newcommand{\eea}{\end{eqnarray}}

\begin{document}
\title{{\bf Revising the predictions of inflation for the cosmic microwave background anisotropies}}
\author{Iván Agulló}\email{ivan.agullo@uv.es}
\affiliation{ {\footnotesize Departamento de Física Teórica and
IFIC, Centro Mixto Universidad de Valencia-CSIC.
    Facultad de Física, Universidad de Valencia,
        Burjassot-46100, Valencia, Spain. }} \affiliation{ {\footnotesize Physics Department, University of
Wisconsin-Milwaukee, P.O.Box 413, Milwaukee, WI 53201 USA}}
\author{José Navarro-Salas}\email{jnavarro@ific.uv.es}
\affiliation{ {\footnotesize Departamento de Física Teórica and
IFIC, Centro Mixto Universidad de Valencia-CSIC.
    Facultad de Física, Universidad de Valencia,
        Burjassot-46100, Valencia, Spain. }}

\author{Gonzalo J. Olmo}\email{olmo@iem.cfmac.csic.es }
\affiliation{\footnotesize Instituto de Estructura de la Materia,
CSIC, Serrano 121, 28006 Madrid, Spain}
\affiliation{ \footnotesize Perimeter Institute for Theoretical
Physics, Waterloo, Ontario, N2L 2Y5 Canada}
\author{Leonard Parker}\email{leonard@uwm.edu}
\affiliation{ {\footnotesize Physics Department, University of
Wisconsin-Milwaukee, P.O.Box 413, Milwaukee, WI 53201 USA}}

\date{July 3th, 2009}

\begin{abstract}
We point out
that if quantum field renormalization is taken into account, and
the  counterterms are evaluated at the Hubble-radius crossing time or  few e-foldings after it, the
predictions of slow-roll inflation for both the scalar and tensorial
power spectrum change significantly. This leads to a
change in the consistency condition that relates the
tensor-to-scalar amplitude ratio with spectral indices.
A reexamination of the potentials $\bf{\phi^2,
\phi^4}$, shows that both are compatible with five-year WMAP data.
Only when the counterterms are evaluated at much larger times
beyond the end of inflation one recovers the standard predictions.
The alternative predictions presented
here may soon come within the range of measurement of
near-future experiments.

\end{abstract}

\pacs{98.80.Cq}

\maketitle
A sufficiently long period of accelerated expansion in the very
early universe is able to solve the questions raised by the standard
big bang cosmology \cite{guth}.
The hot big bang cosmology is an
extremely successful theory. It explains the existence of the cosmic
microwave background (CMB) and its thermal nature, the observed
expansion of the universe, the abundance of light elements and the
astrophysical fits for the age of the universe. However, it leaves
without answer why our universe appears so homogeneous and nearly
flat at large scales. Inflation offers a natural answer to these
questions and, at the same time, provides a predictive mechanism to
account for the small observed inhomogeneities
\cite{mukhanov-chibisov}
responsible for the structure formation
in the universe and the anisotropies present in the cosmic microwave
background (CMB), as first detected by the COBE satellite  and
further analyzed by the Wilkinson Microwave Anisotropy Probe (WMAP)
satellite \cite{WMAP5}. Inflation predicts production of primordial
density perturbations and relic gravitational waves as
amplifications of  vacuum fluctuations together with a
quantum-to-classical transition at the scale of Hubble sphere crossing.
Primordial perturbations
leave an imprint in the CMB anisotropies, which are, therefore, of
major  importance for understanding our universe and its origin. The
potential-energy density of  a scalar (inflaton) field is assumed to
cause the inflationary  expansion, and the amplification of its
quantum fluctuations and those of the metric are inevitable
consequences in an expanding universe \cite{parker69}. The metric
fluctuations provide the initial conditions for the acoustic
oscillations of the plasma at the onset of the subsequent
radiation-dominated epoch. The detection of the effects of primordial gravitational
waves in future high-precision measurements of the CMB anisotropies,
as for instance in the PLANCK satellite mission \cite{planck}, will
serve as a highly non-trivial test for inflation. Therefore, it is
particularly important to scrutinize, from all points of view, the
standard predictions of inflation (as summarized for instance, in
\cite{LiddleLyth2000})  to be tested
empirically. This is the aim of this paper. We point out that  if
quantum field renormalization is taken into account, as in the
experimentally tested Casimir effect, the quantitative predictions
of inflation change significantly,
and may be tested in forthcoming
CMB measurements.

The scalar perturbations, which constitute the  ``seeds'' for
structure formation, are characterized by the power spectrum \be
\label{sps} P_{\cal R}(k) = \left (\frac{H}{\dot{\phi}}\right )^2
\left (\frac{H}{2\pi}\right )^2  \ , \ee where $\phi$ represents
the inflaton scalar field, which dominates the energy density
during inflation. Here $H$ stands for the Hubble rate $H\equiv
\dot a/a$, where $a(t)$ is the expansion factor and dot means
derivative with respect to the comoving time. The above expression
is evaluated at the Hubble radius crossing time $t_k$ (usually called ``horizon crossing'' time), where $k/a(t_k)=H$.
In the typical slow-roll inflationary scenario the homogeneous
part of the inflaton field $\phi_0(t)$ rolls slowly down its
potential $V(\phi)$ towards a minimum. Both $\phi$ and $H\equiv
\sqrt{\frac{8\pi G}{3} V(\phi_0)}$ are changing very gradually and
this change is parameterized by the slow-roll parameters
$\epsilon, \eta$, where $ \epsilon \equiv -\dot H/H^2$, and $\eta
-\epsilon \equiv \ddot \phi/(H\dot \phi)$. These parameters can be
related to the derivatives of the inflaton potential $ \epsilon =
(M_P^2/2)(V'/V)^2, \eta= M_P^2(V''/V)$, where $M_P=1/\sqrt{8\pi
G}$ is the reduced Planck mass in natural units $\hbar=1=c$. In
the slow-roll approximation, $\epsilon \ll 1$ and $|\eta| \ll 1$,
the scalar power spectrum turns out to be \be \label{psR}P_{\cal
R}(k) = \frac{1}{2M_P^2\epsilon (t_k)}\left
(\frac{H(t_k)}{2\pi}\right )^2 \ . \ee  In addition, the power
spectrum of tensor fluctuations is given by \be \label{psT}P_t(k)
= \frac{8}{M_P^2} \left (\frac{H(t_k)}{2\pi}\right )^2\ , \ee and
the tensor-to-scalar ratio is then $r=P_t/P_{\cal R}= 16\epsilon$.
The power spectra are not exactly scale invariant, they can vary
with $k$ and this dependence is parameterized by the
 scalar and tensorial  spectral
indices $n_s-1\equiv  d \ln P_{\cal R}/d\ln k, \ n_t \equiv d \ln
P_{t}/d\ln k$. Since these indices are  related to the slow-roll
parameters $n_s-1=  -6\epsilon + 2\eta\ ,  n_t = -2\epsilon$, one
generates immediately  the  consistency relation $r= -8n_t$, which
should be verified by any single-field slow-roll inflationary
model, irrespective of the particular form of the potential.

In this paper we shall reexamine, on the basis of general
principles of quantum field theory in an expanding background
\cite{parker-toms, birrel-davies}, the fundamental expressions
(\ref{psR}-\ref{psT}) for the scalar and tensorial power spectrum.
In doing this we shall also be led to modify the expressions  for
the spectral indices in terms of the slow-roll parameters and,
therefore, to generate a new consistency relation. We will have
all the necessary ingredients to reexamine the observational
predictions of inflationary models and here we shall do that for
some of the most significant  models.

The second factor in the  fundamental relation (\ref{sps}) has its
origin in the quantum fluctuation  of the scalar inflaton field
$\phi$. The  first-order perturbation $\delta \phi$, where $\phi =
\phi_0(t) + \delta \phi (x)$, obeys the wave equation \be
\label{weq}\partial_t^2 \delta \phi + 3H \partial_t \delta \phi
-a^{-2}\sum_{i=1}^3\partial_i^2 \delta \phi + m^2 \delta \phi =0 \
, \ee where $a(t)$ is the expansion factor of the unperturbed
homogeneous and spatially flat metric $ds^2 = -dt^2 +
a^2(t)d\vec{x}^2$.
 The effective mass term, which is necessarily small in the slow-roll approximation, is given by the second derivative of the potential:
  $m^2= V''(\phi_0)$. Moreover, the fundamental relation (\ref{psT}) has also the same quantum origin. The two independent polarizations
  of tensorial modes can be described by a couple of scalar fields $h_{+, \times}$ obeying the above wave equation with zero mass.
  The relation between $h$ (we omit the subindex $+$ or $\times$) and $\delta \phi$ is given by  $\delta \phi \equiv  h/\sqrt{16\pi G}$.

  Let us focus
  first on these tensorial perturbations. The form of the modes is  then $h_{\vec{k}}(t,
\vec{x}) = (-16\pi G\tau
\pi/4(2\pi)^3a^2)^{1/2}H^{(1)}_{\nu}(-k\tau)e^{i\vec{k}\vec{x}}$.
The index of the Bessel function is  $\nu=\sqrt{9/4 +3\epsilon}$
and the so-called conformal time, $\tau\equiv\int dt/a(t)$, is
given by $\tau=-(1+\epsilon)/aH$. Notice that, at early times, the
amplitude of oscillations depends on $k$ in a way similar to that
of a massless field in Minkowski space. As time evolves, the
comoving wavelength reaches the Hubble horizon length at $t_k$.
  A few Hubble times after horizon exit the
amplitude freezes completely and, for all subsequent times t, one
has the constant value $|h_{\vec k}|^2 = \frac{
GH^2(t_k)}{\pi^2k^3}$. Because of the loss of phase information, the modes of the perturbations
soon take on classical properties \cite{KieferPointerStates2007}. The
freezing amplitude is usually codified through the quantity
$\Delta_h^2(k,t)$, defined in general by $\Delta_h^2= 4\pi k^3
|h_{\vec k}(t)|^2$, and  evaluated at the horizon crossing $t_k$
(or a few Hubble times after it). Taking into account the two
polarizations,   one finally obtains a nearly ``scale free''
tensorial power spectrum (\ref{psT}). In a similar way one obtains
the scalar power spectrum (\ref{psR}).

 At this stage it is important to remark that the above definition of
the power spectrum is such that the quantum fluctuations of the
perturbations in position space satisfy the relation \be \langle
h^2(\vec{x}, t) \rangle = \int_0^{\infty}\frac{dk}{k}\Delta^2_h
(k,t) \ . \label{eq:variance}\ee This quantity represents the
variance of the Gaussian probability distribution associated to
$h(\vec{x}, t)$.
 However, in the form given by (\ref{eq:variance}), it is divergent. It might be argued, as is common when dealing with random fields,
 that this divergence can be eliminated by smoothing out the field on a certain scale $R$ to remove the Fourier modes
 with $k^{-1}<R$.  It can also be argued that the value of $\langle h^2 \rangle$ is unimportant and regard the (finite) two-point
 function $\langle h(x_1)h(x_2) \rangle$, uniquely defined by $\Delta^2_h (k)$, as the basic object.
However, it is our view to regard the variance as the basic
physical object, which defines the amplitude of fluctuations in
position space,  and treat $h(\vec{x}, t)$ as a  quantum entity.
Therefore, if one regards $h(\vec{x}, t)$ as a quantum field, and
$ \langle h^2(\vec{x}, t) \rangle$ its
 two-point function at coincident points, the divergences must be handled in a different way.
  Taking into account the large $k$ asymptotic form of the modes one can estimate the form of the integral at an arbitrary time
\be \langle h^2(\vec{x}, t) \rangle \approx
\int_0^{\infty}\frac{dk}{k} \frac{16\pi G k^3}{4\pi^2a^3}\left [
\frac{a}{k}[1 +\frac{(2+3\epsilon)}{2k^2\tau^2}] + ... \right]\ .
\ee The first term produces a quadratic divergence, which is the
typical singularity of a quantum field in Minkowski space. The
second term produces a logarithmic divergence, which is typical of
a massless field in an expanding universe. For a quantum
mechanical system, with a finite number
 of degrees of freedom, such a divergent behavior would not arise and one need not worry about the definition
 of the physical
 power spectrum. However, a quantum field is neither a random field nor a quantum mechanical system with a finite number of degrees of freedom.
 The existence of divergences  tells us that the nature of a quantum field is more involved.
If we wish to  have a finite expression for $\langle h^2(\vec{x},
t) \rangle$ one should subtract the divergences in a consistent
way.
  In technical words, one should renormalize  the quantum field.
Since the physically relevant quantity (power spectrum) is
expressed in momentum space, the natural renormalization scheme to
apply is the so-called adiabatic subtraction \cite{parker07}, as
it renormalizes the theory in momentum space. Adiabatic
renormalization \cite{parker-fulling, parker-toms, birrel-davies}
removes the divergences present in the formal expression
(\ref{eq:variance}) by subtracting $(16\pi G k^3/4\pi^2
a^3)[1/w+\dot{a}^2/2a^2w^3 +\ddot{a}/2aw^3]$ mode by mode in the
integral (\ref{eq:variance}), where $w$ is the frequency of the
modes.  The subtraction of the first term $(16\pi G k^3/4\pi^2 a^3
w)$ to cancel the typical flat space vacuum fluctuations was
already considered in \cite{booklinde}.
However, the additional terms, proportional to $\dot{a}^2$ and
$\ddot{a}$ are necessary to properly perform the renormalization
in an expanding universe. Subtracting consistently the divergent
terms one obtains the finite expression \bea \langle h^2(\vec{x},
t) \rangle &=& \int_0^{\infty}\frac{dk}{k} \frac{16\pi G
k^3}{4\pi^2a^3} [-a\frac{\tau \pi}{2}|H^{(1)}_{\nu}(-k\tau)|^2
\nonumber \\ &-& \frac{a}{k}[1
+\frac{(2+3\epsilon)}{2k^2\tau^2}]]\ . \eea For the idealized case
of a strictly constant $H$ the subtraction exactly cancels out the
vacuum amplitude \cite{parker07} since $H^{(1)}_{3/2} (x) =  i \exp(i x) \sqrt{2/ \pi x} (-1 + i/x) $. From this it follows
that in pure de Sitter inflation there is no production of
gravitational waves. One would thus expect that the tensorial
power spectrum were proportional to the $\epsilon$ slow-roll
parameter, which parameterizes the slow change in $H(t)$. The fluctuations still acquire classical properties through decoherence in a time of order $H^{-1}$ after their wavelengths become larger than the Hubble radius during inflation \cite{KieferPointerStates2007}. Therefore, it is natural
to evaluate the power spectrum (with the corresponding adiabatic
counterterms) a few Hubble times after the time $t_k$. Since the
results will not be far different from those at $t_k$, we use that
time to characterize the results. We comment further on this point after (\ref{r}). Evaluation of the above integral
at the time $t_k$ leads to the redefinition \be \Delta^2_{h}(k) =
4G H^2\left
[\frac{1+\epsilon}{2}|H^{(1)}_{\nu}(-k\tau=1+\epsilon)|^2-
 \frac{(4-\epsilon)}{2\pi} \right ]\ , \ee which turns out to be proportional to $\epsilon$, $
\Delta^2_{h}(k)= \alpha 16\pi G(H(t_k)/2\pi)^2\epsilon (t_k) + O(\epsilon^2)$,
where $\alpha$ is a numerical coefficient of order unity, $\alpha
\approx 0.904$. Taking into account the  usual conventions, we get, at
leading order in $\epsilon$, \be \label{psT2} P_t(k)=
 \frac{8 \alpha}{M_P^2}\left (\frac{H(t_k)}{2\pi} \right )^2 \epsilon(t_k) \ . \ee
One can proceed in the same way to reevaluate the scalar power
spectrum. However, to be precise in the calculation, one must keep
the mass term in (\ref{weq}) and also take into account the slow
decay of the Hubble rate $H$, as above. The latter is controlled
by the slow-roll parameter $\epsilon$, while the former is
captured by the slowly changing parameter $\eta \equiv
M_P^2(V''/V)$, which gives $m^2 =3 \eta H^2$. The form of the
scalar modes is the same as that for the tensorial ones, up to the
fact that the coefficient $\sqrt{16\pi G}$ is now absent and the
Bessel index is now $\nu^2 = 9/4 +3\epsilon -3\eta$. The adiabatic
subtracting terms are then those already present for a massless
field, with $w(t) = \sqrt{k^2/a^2 + m^2}$, together with new terms
proportional to the mass $(k^3/4\pi^2
a^3)[m^2\dot{a}^2/2a^2w^5+m^2\ddot{a}/4aw^5
-5m^4\dot{a}^2/8a^2w^7]$. Therefore, after renormalization, the
final result for the  scalar power spectrum is \be
\label{sps2}P_{\cal R}=
\frac{1}{2M_p^2\epsilon(t_k)}\left(\frac{H(t_k)}{2\pi}\right)^2(\alpha\epsilon(t_k)
+3\beta{\eta}(t_k)) \ , \ee where $\beta \approx 0.448$ is another
numerical coefficient. In the case of eternal de Sitter expansion,
one recovers the previous results of
\cite{parker07,agullo-navarro-salas-olmo-parker}.

The above expression, together with (\ref{psT2}) for the tensorial
power spectrum, constitute our main results and lead to a change
in the testable  predictions of  inflation. To see this we have to
consider  the scalar and tensorial spectral indices.
The
standard expression for them, in terms of the slow-roll parameters,
is $n_s-1= -6\epsilon + 2\eta\ , n_t = -2\epsilon$. However, if we
invoke renormalization we get \be \label{nt2}n_t=
2(\epsilon-\eta)\ee \be \label{ns2}n_s-1 = -6\epsilon+2 \eta +
\frac{4\alpha\epsilon^2+(6\beta-2\alpha)\epsilon \eta - 3\beta
(n'_t/2 -4\epsilon(\epsilon -\eta))}{\alpha \epsilon + 3\beta \eta }
\ee where $n'_t$ is the running of the tensorial index $n'_t\equiv
d{n_t}/d\ln{k}$. Note that $n'_t$ can be expressed in terms of the
slow-roll parameters as $n'_t=8 \epsilon( \epsilon-\eta) + 2 \xi$,
where $\xi$ is defined by $\xi\equiv M_P^4 (V' V'''/V^2)$. The above
formulae provide, implicitly, algebraic relations between physical
observables. The most important one is the relation between the
tensor-to-scalar ratio $r\equiv P_t/P_{\cal {R}}$ with the spectral
and running indices: $r=r(n_t, n_s, n'_t)$. In the
simplest case of $n'_t=0$, and taking the approximation $\alpha
\approx 2 \beta$, the new consistency condition becomes \be
r=1-n_s+\frac{96}{25}n_t+\frac{11}{5}\sqrt{(1-n_s)^2+\frac{96}{25}
n_t^2}\ . \ee
\begin{figure}[htbp]
\begin{center}
\includegraphics[width=5.7cm]{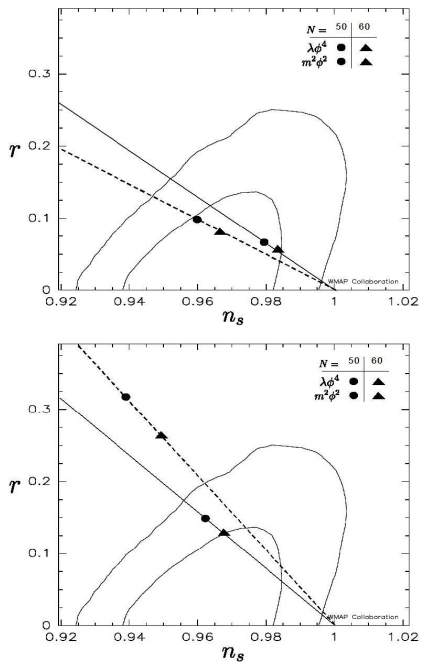}
\caption{Plot of $r$ versus $n_s$. The contours show the $68\%$
and $95\%$ CL derived from WMAP5 (in combination with BAO$+$SN)
\cite{WMAP5}. We consider
$V(\phi)= m^2 \phi^2$ (solid line), $V(\phi)= \lambda \phi^4$
(dashed line). The symbols show the prediction from each of this
models in terms of the number $N$ of e-folds. The top part corresponds to the prediction of
our formulae, while the bottom one corresponds to the standard
prediction.}
\end{center}
\end{figure}
This relation could be potentially checked, and compared with the
standard one $r= -8n_t$, in near future high-precision anisotropy
measurements of the CMB. However, we can already  contrast,
partially,  our predictions with the standard ones on the basis of
the five year WMAP results by performing a model by model
analysis. As a representative example we shall reexamine  the
monomial chaotic potential  $V(\phi)= \lambda M_P^{(4-p)} \phi^p$
and compare with standard theoretical and experimental results
\cite{WMAP5}.  It is not difficult to obtain that \be \label{r}
r=\frac{\alpha p^2}{(3\beta(p-1)/2+\alpha p/4)N}  \ \ , \ \ 1-
n_s= \frac{p}{2N}, \ee where $N\equiv \ln{a_{end}/a_{WMAP}}$ is the
number of $e$-folds of inflation between the end of inflation and
the epoch when the wavelength of fluctuations that WMAP detects
left the horizon. If
the adiabatic counterterms were evaluated  some  $e$-folds $n$ after $t_k$ but still well before the end of inflation ($n\epsilon \ll 1$), then  the value of $\alpha$ in the tensorial power spectrum would change  to $\alpha \approx 2n$. A similar calculation for the scalar power spectrum should also be carried out. This computation should deal with the gauge-invariant quantity ${\cal{R}}_{k}$, which is conserved outside the Hubble sphere and satisfies the same equation as $h_{\vec{k}}$, up to the replacement $a \to a\dot{\phi}_0/H$. The result is then $\alpha \approx 6n$, $3\beta \approx -2n$.
One then finds that the spectral index $n_s$ in (\ref{r}) is unchanged, and the ratio $r$ becomes $r= 4p^2/(p+2)N$, which is not sensitive to the value of $n$.
If the counterterms are evaluated far beyond the end of inflation, one recovers the standard predictions
 $r=4p/N$ and
$1-n_s=(p+2)/2N$. However, since  primordial fluctuations acquire classical properties through decoherence when their wavelengths become larger than the Hubble radius \cite{KieferPointerStates2007}, we find it natural to evaluate the spectra at a time close to  $t_k$, which implies significant deviations from the standard prediction (see (\ref{r})). Note that, with the present understanding of the nonlinear aspects of quantum gravity, it is difficult to reach a definitive conclusion regarding this question, so the fact that there are observable differences offers a deep way to experimentally probe this question.
In fact, if we  compare the results at $t_k$ with the WMAP
5-year data for the representative values $p=2$ and $p=4$ (Figure
1) we find that  both models are compatible with the experimental data for the
reasonable range of $N$ between $50$ and $60$ (Figure 1 top). This is in
sharp contrast with the prediction of the standard approach
(Figure 1 bottom) where the monomial potential with $p=4$ is
excluded convincingly. With the new predictions both models
are compatible with WMAP data for all
the values of $N$ between $50$ and $60$. This is also true even if the counterterms are evaluated $n$ $e$-folds after $t_k$, with $n\epsilon \ll 1$.
The alternative
predictions presented here may soon come within the range of
measurement of high-precision CMB experiments, as expected in the
PLANCK satellite mission.
\noindent { \bf Acknowledgements} This
work has been supported by grant FIS2008-06078-C03-02. L.P. has been partly
supported by NSF grants PHY-0071044 and PHY-0503366 and by
a UWM RGI grant. I.A. and G.O.
thank MICINN for a FPU grant and a JdC contract, respectively.
\\

\end{document}